# A SEGMENTATION-ORIENTED INTER-CLASS TRANSFER METHOD: APPLICATION TO RETINAL VESSEL SEGMENTATION

*Chengzhi Shi[1,2], Jihong Liu[1]*, Dali Chen[1]*

[1]College of Information Science and Engineering, Northeastern University, Shenyang, 110819, Liaoning, China
[2] Department of Computing, Imperial College London, SW7 2AZ, London, UK
*Corresponding author, Email: Liujihong@ise.neu.edu.cn

## ABSTRACT

Retinal vessel segmentation, as a principal nonintrusive diagnose method for ophthalmology diseases or diabetics, suffers from data scarcity due to requiring pixel-wise labels. In this paper, we proposed a convenient patch-based two-stage transfer method. First, based on the information bottleneck theory, we insert one dimensionality-reduced layer for task-specific feature space. Next, the semi-supervised clustering is conducted to select instances, from different sources databases, possessing similarities in the feature space. Surprisingly, we empirically demonstrate that images from different classes possessing similarities contribute to better performance than some same-class instances. The proposed framework achieved an accuracy of 97%, 96.8%, and 96.77% on DRIVE, STARE, and HRF respectively, outperforming current methods and independent human observers (DRIVE (96.37%) and STARE (93.39%)).

## 1. INTRODUCTION

After being processed, retinal fundus images are used to diagnose diseases like diabetic retinopathy, glaucoma, cardiovascular and hypertension. Accurately extracting vessels is still a difficult task for several reasons like the presence of noise, the low contrast between vasculature and background, the variations in illumination and shape. Moreover, due to the presence of lesions, exudates, and other pathological effects, the image may have large abnormal regions [1]. Therefore, to diagnose and grade the diseases, an effective vessel segmentation method is indispensable.

However, annotated medical segmentation datasets, especially in the specific field, are still relatively unavailable.

Consequently, semi-supervised methods, which require small annotated datasets, are proposed to relieve the pressure in medical image processing [3].

Simultaneously, transfer learning methods are applied to image processing. Varga et al. introduce a similarity measure between images to help label transfer with low-level visual descriptors[4]. In terms of biomedical imaging, Li et al. proposed transfer learning with pre-trained models in diabetic retinopathy disease classification [5]. Zheng proposed cross-modality transfer learning for shape priors, applying information from CT to MRI, in medical imaging [6]. In [7], Wu et al. applied constrained deep transfer learning combined with feature learning based on Deep Boltzmann Machines (DBM) to realize iteratively transfer satisfying specific priori conditions like eye shapes.

In our paper, we propose a two-stage method to solve annotated data shortage in retinal fundus vessel segmentation. Our method, iteratively seeking proper feature space to achieve most task-specific transfer, is introduced to transfer data from other organs (e.g. vessels in lungs) or from different contexts (e.g. neurons) to solve overfitting and facilitate to learn features

## 2. PREPROCESSING AND METHODOLOGY

### 2.1. Preprocessing

We work on the green channel of the retinal image since the green channel of the RGB color retinal image presents a higher contrast between the vessels and the background. In addition, we normalize the green channels of images by using the following formula:

$$I_{PG} = \frac{I_G - \mu}{\sigma} \tag{1}$$

where μ and σ denote the mean and standard deviation of the

~~Thanks for Natural Science Foundation~~ of China (No.61773104).

image data. $I_{PG}$ and $I_G$ are the preprocessed and original green channel input, respectively.

We apply the CLAHE [8] operator to obtain a local contrast enhanced retinal image. Then, Gamma adjustment [9] is applied to normalized images after contrast limited adaptive histogram equalization. The gamma adjustment is mainly used for image correction, and the image with too high grayscale or low grayscale is corrected to enhance the contrast. On DRIVE, we apply a gamma value of 1.2.

## 2.2. General framework

The method will be introduced regarding DRIVE as a task database. The framework as shown in Fig. 1 takes cropped patches as input. Cropping not only augments training data but also converts segment object from entire retinal images to small patches including plenty of thin vessels.

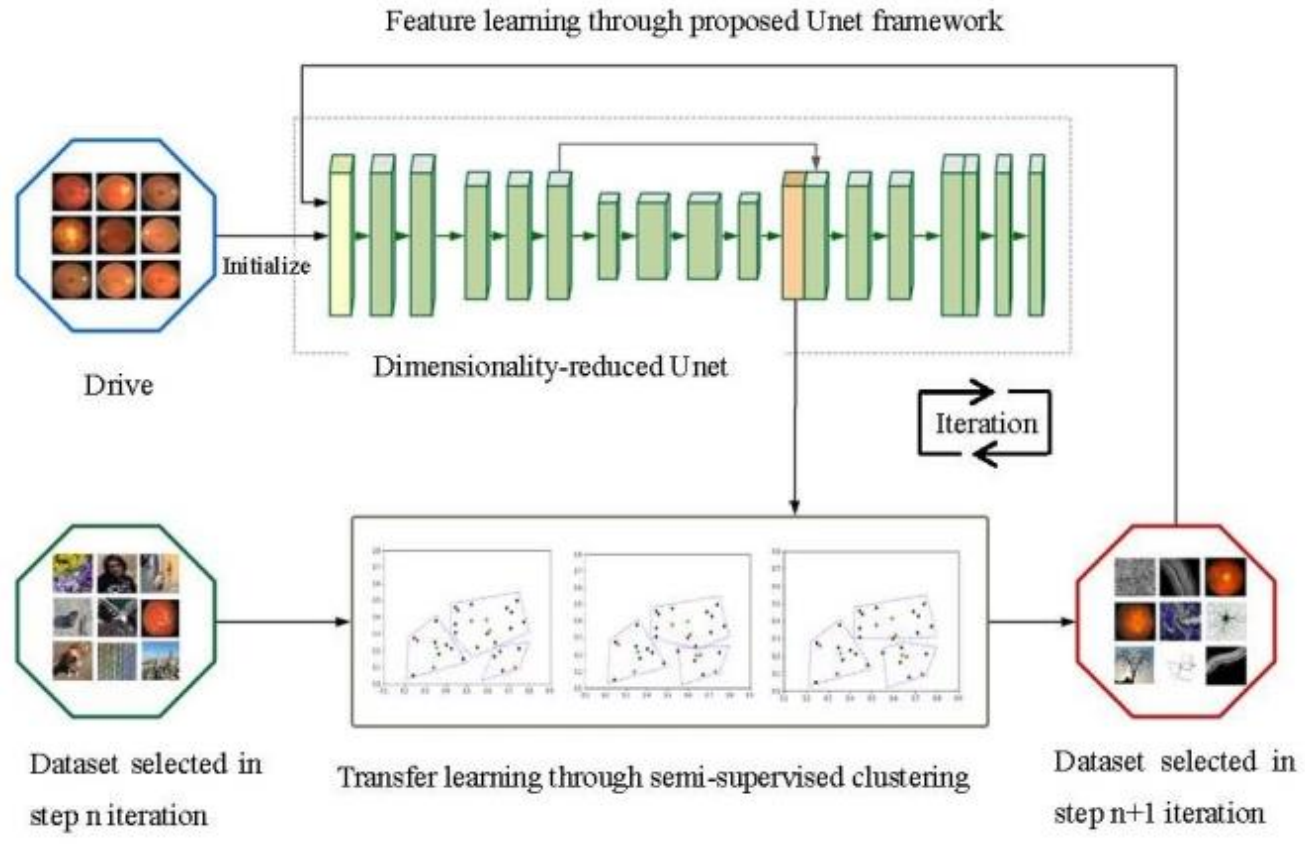

Fig. 1 The overview of the proposed general framework. It consists of dimensionality-reduced Unet to establish a feature space from cropped images and Semi-Supervised Clustering algorithm with a voting mechanism to select effective transfer samples.

Then, to obtain the result of a complete image, we must stitch predicted patches into a complete image. This strategy utilizes context information and helps eliminate accidental errors.

Another key point in our proposed framework is selectively transferring instances for segmentation. Feature space is one of the most important factors in transfer learning, while our method is based on datasets from various context.

Therefore, iterations between transfer learning and feature learning are needed when seeking a proper latent feature space to narrow the gap between the source domain and target domain to describes both DRIVE and other datasets better.

## 2.3. Feature learning through dimensionality-reduced Unet

The intuition behind the Dimensionality-Reduced Layer is the Information Bottleneck Theory. The proposed neural network architecture is derived from the Unet architecture, while the database of retinal fundus vessels is quite limited, and features are relatively less diversified.

In the general supervised setting, we want to learn the conditional distribution $p(\boldsymbol{y}|\boldsymbol{x})$ of some random variable $\mathbf{y}$, which we refer to as the task, given (samples of the) input data $\mathbf{x}$. In our case, $\mathbf{x}$ can also refer to the output of hidden layers, which is mostly more than 20,000 dimensions like the layer in the middle of Fig. 2. In such cases, a large part of the variability in $\mathbf{x}$ is actually due to nuisance factors that affect the data, but are otherwise irrelevant for the task.

In terms of representation learning, the output of the middle layers is latent variables $\mathbf{z}$ that can describe $\mathbf{x}$ with much fewer dimensions. Thus, the problem can be formulated as

$$\begin{aligned} &\text{minimize} \quad I(\mathbf{x},\mathbf{z}) \\ &\text{s.t.} \quad I(\mathbf{x},\mathbf{y}) - I(\mathbf{z},\mathbf{y}) = 0 \end{aligned} \tag{2}$$

where $I(\mathbf{x},\mathbf{z})$ refers to the mutual information between $\mathbf{x}$ and $\mathbf{z}$. Thus, this optimization problems means, to improve our specific task $\boldsymbol{y}$, irrelevant latent variables must be ruled out, while the most relevant latent variables should be maintained.

Since the problem is hard to solve, [10] propose a generalization known as the Information Bottleneck Principle, associated with the Lagrangian form as

$$\mathcal{L} = I(\mathbf{x},\mathbf{z}) + \lambda(I(\mathbf{x},\mathbf{y}) - I(\mathbf{z},\mathbf{y})) \tag{3}$$

And this problem would be solved during the training of Dimensionality-Reduced Network. Based on this solution, we can obtain the optimal transformation corresponds to our specific task, retinal vessel segmentation. The solution and transformation are

$$\begin{aligned} \frac{\partial \mathcal{L}(\mathbf{z}^*)}{\partial \mathbf{z}} &= 0 \\ \mathbf{z}^* &= f_\theta(\mathbf{x}) \\ i^* \in \arg\min_i f_\theta(\mathbf{x}^{Target}) &- f_\theta(\mathbf{x}_i^{Source}) \end{aligned} \tag{4}$$

where θ is the weights, while $i$ and $i^*$ refer to the index of images from the source domain and the image to transfer, respectively. In other words, in this latent space, features irrelevant to vessel segmentation $(\boldsymbol{y})$ will be ignored.

In terms of structures, our network is simplified based on U-net. The architecture is shown in Fig. 2. The loss function is the cross-entropy and the Adam optimizer is employed [10]. The dimensionality-reduced layer which implements the feature extraction after dimensionality reduction is framed.

## 2.4. Transfer through semi-supervised clustering

With learned feature space, Semi-supervised K-Means method is used to selectively transfer similar samples to facilitate feature learning of Unet in progress or decide transferred images in the end. Supervisory information is a small amount of picture-level labeled samples (i.e. all images in DRIVE and images selected from CHASE, STARE and HRF). The sample set is defined as $D = \{x^1, x^2, \ldots, x^m\}$.

Since we crop the input as we proposed, each of the images in the sample set whether labeled or unlabeled ought to be cropped. We select ten patches from each image so that the sample set is redefined as $D' = \{x_1^1, x_2^1, \dots, x_{10}^1, \dots, x_{10}^m\}$.

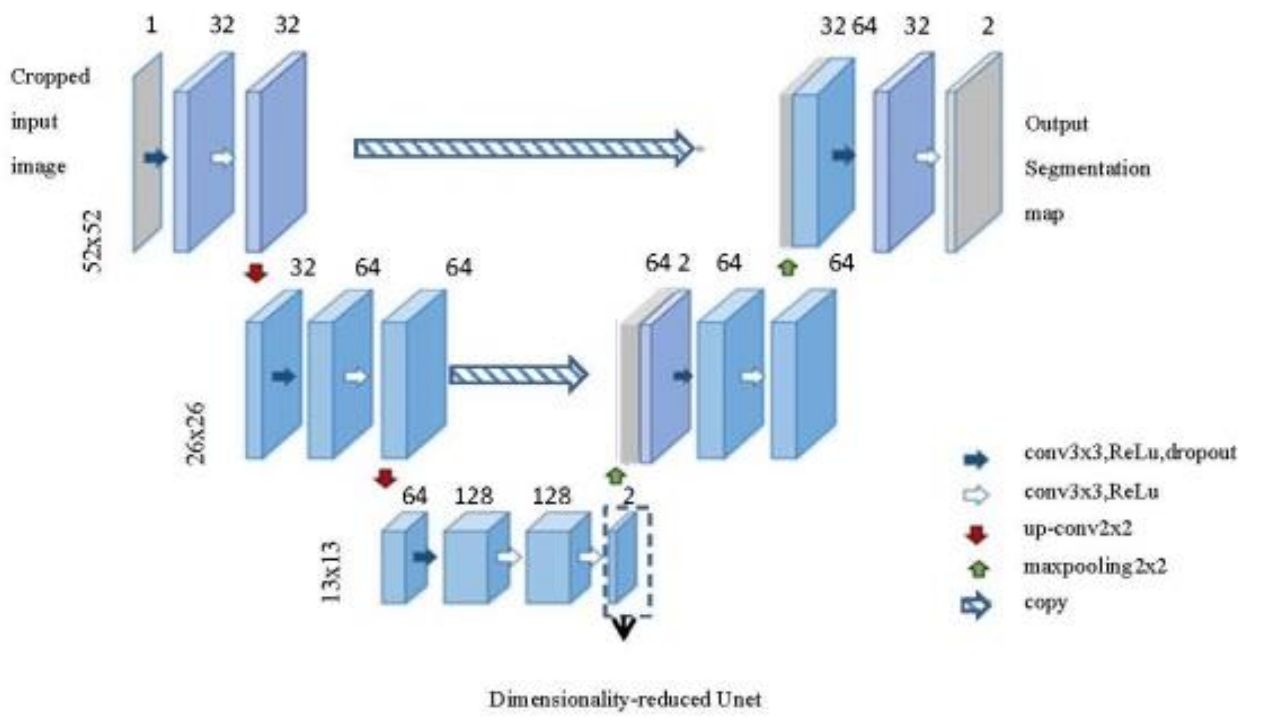

Fig. 2. Dimensionality-reduced Unet architecture. Blue boxes correspond to multi-channel feature maps. The dimensions are around the box, with arrows denoting different operations. Besides, to realize rapid iterations and guarantee the convergence of clustering, feature maps should be reduced with accuracy maintained.

The labeled sample set is defined as $S = \cup_{j=1}^{k} S_j \subset D'$, where $S_j \neq \emptyset$ is the sample of the j[th] cluster. Labeled samples are used as seeds directly and to initialize the k cluster centers of the K-Means algorithm. In addition, the cluster membership of the seed samples does not change during the cluster's iterative updating. The Iterative update stops when the mean vector of cluster centers become unchanged.

Based on the Constrained Seed K-Means algorithm described above, a patch-based voting mechanism is introduced. The pseudocode of our algorithm is described as following.

**Algorithm: Semi-Supervised Clustering**

---

**Input:** Dataset: $D = \{x^1, x^2, \dots, x^m\}$;
Each sample $x$ is cropped randomly into 10 patches: $D' = \{x_1^1, x_2^1, \dots, x_{10}^1, \dots, x_{10}^m\}$;
Labeled dataset: $S = \cup_{j=1}^{k} S_j \subset D'$
($S_0$:DRIVE,$S_1$:STARE,$S_2$:HRF);
Number of clusters: k(k=3)

**Process:**

**for** j=1, 2, ..., k **do**
  Initialize mean vector $\mu_j = \frac{1}{|S_j|}\sum_{x \in S_j} x$
**end for**
**repeat**
  Initialize set of clusters $C_j \leftarrow \emptyset (1 \leq j \leq k)$
  **for** j =1, ..., k **do**
    **for all** x ∈ $S_j$ **do**
      First divide labeled samples into corresponding cluster collection $C_j \leftarrow C_j \cup \{x\}$
    **end for**
  **end for**
  **for all** $x^i \epsilon D \backslash S$ (1 ≤ i ≤ m) **do**
    Initialize set of cluster results of patches of each group $r_N \leftarrow \emptyset$
    **for all** $x_n^i \epsilon D \backslash S$ (1 ≤ n ≤ 10) **do**
      Compute distance $d_{nj}^i = \|x_n^i - \mu_j\|_2$
      Find nearest cluster to sample $x_n^i$: $r_n = \arg min_{j \in \{1,2,..,k\}} d_{nj}^i$
      Put every individual clustered result into a collection $r_N \leftarrow r_N \cup \{r_n\}$
    **end for**
    Find the element most often appear in $r_N$ : r
    Divide $x^i$ into the corresponding cluster collection $C_r \leftarrow C_r \cup \{x^i\}$
  **end for**
  **for** j=1, 2, ..., k **do**
    Update mean vector $\mu_j = \frac{1}{|C_j|}\sum_{x \in C_j} x$
  **end for**
**until** mean vector does not update

---

## 3. EXPERIMENT AND RESULTS

### 3.1. Databases

In the experimental part, we mainly rely on the public dataset DRIVE [11] for evaluating the feasibility and validity of the proposed algorithm.

Due to the small sample size of DRIVE, we are motivated to use several datasets for transfer, either because they (STARE [12], CHASE [13], HRF [14]) are also fundus image datasets (same-class), or because their (CREMI [15], STARERAW [16], VascuSynth [17], Cellular_2DNuclei [18]) domain may have commonality with the source domain of DRVIE (cross-class datasets, but with filament structures, medical-related and textures etc.)

### 3.2. Training parameters

Here, we would introduce our strategies, regarding DRIVE as the task dataset as well. Training consists of iterations between feature learning through Unet whose input images are cropped. Every image is cropped into 500 52x52 patches and selectively transferred through semi-supervised clustering.

Unet provides a dimensionality-reduced task-specific feature space describing images for semi-supervised clustering. In return, selected images and DRIVE help Unet to build a feature space with less gap between the source domain and target domain. The iteration ends when the performance reaches the optimal point and degenerates.

In Unet, we initialize weights by Xavier method [19] and train them applying Adam optimization [10].

In semi-supervised clustering, we use datasets HRF and STARE, which are also retinal images and tested to be helpful, with DRIVE, as seeds (labeled sets) to guide clustering images from other datasets. Correspondingly, when evaluating STARE, we take DRIVE and HRF as guidance seeds (only training sets).

To calculate loss, we must stitch predicted patches into a complete image. Here we take the advantages of overlapped patches, which means the segment result of each pixel is determined by all patches covering that pixel.

### 3.3. Validation and experiments

Based on our framework introduced in Section III, we firstly conduct experiments regarding DRIVE as the target task and other datasets as source tasks.

Our method is also tested on official STARE and HRF test sets respectively. We present the performance of networks tested on the test sets in terms of area under the ROC curve (AUC), accuracy (Acc), sensitivity (Sen) and specificity (Spe). And the area under the ROC curve (AUC) is calculated for quality evaluation.

### 3.3.1. Validation experiments on DRIVE

To validate the proposed method, the result is compared with other segmentation methods: (1) the original patch-based Unet without transfer learning [19]; independent human observers (here, the second official manual segmentation result of DRIVE is regarded as the performance of a human observer) [11]; the no-pool architecture in knowledge transfer mode, which is directly adding the training set of STARE to DRIVE [20].

As is shown in Table II, the proposed method outperforms the independent human observer and the original patch-based Unet. Moreover, through Table 1, we can conclude that the knowledge transfer, combining STARE and DRIVE, used in [20], simply combing similar databases is not effective.

**Table 1** Comparison between Manual Segment, Patch-Based Unet and Proposed Patch-Based Transfer Method on Drive

| Method | Transfer | Acc | Sen | Spe | AUC |
|---|---|---|---|---|---|
| *Human observer* | - | 0.9637 | 0.7742 | 0.9819 | - |
| *No-Pool Architecture1[20]* | ✓ | 0.9491 | - | - | 0.9700 |
| *No-Pool Architecture2[20]* | × | 0.9495 | - | - | 0.9720 |
| *Feng [21]* | × | 0.9560 | 0.7811 | 0.9839 | 0.9792 |
| *Proposed Method* | ✓ | **0.9700** | **0.7864** | **0.9876** | **0.9862** |

### 3.3.2. Comparison with proposed methods

To visualize the performance, comparisons between the manual segmentation and predicted result are shown in Fig. 3.

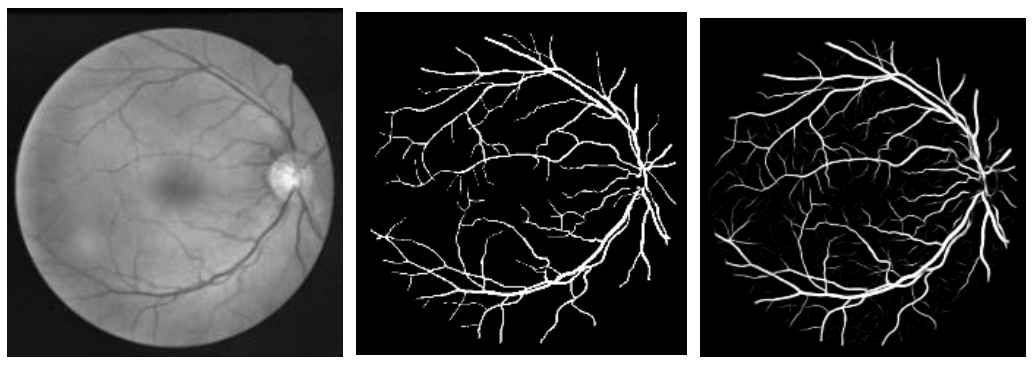

(a) (b) (c)

Fig. 3. Exemplar vessel segmentation results of the $20^{th}$ image in test set of DRIVE. (a) the fundus image after preprocessing, (b) the manual annotation and (c) the predicted probability map.

As shown in Table 2, our approach achieves state-of-the-art performance, because our method managed to accurately segment both thick and thin vessels without noise points.

With the same architecture applied to DRIVE, we test the network on STARE and HRF. The results are shown in Table. 3 and Table 4. After transfer and threshold operation, the networks' performance reaches a state-of-the-art level as well, with the help of several other databases.

Among those datasets, except for retinal databases, datasets like CREMI (a neuron dataset) are tested to be most effective on promoting retinal vessel segments. However, STARERAW, which is also a retinal image database, brings negative effects to retinal vessel segmentation, due to its ophthalmology diseases.

**Table 2** Comparison results on the DRIVE datasets

| Segment method | Acc | Sen | Spe | AUC |
|---|---|---|---|---|
| *Human observer* | 0.9637 | 0.7742 | 0.9819 | 0.8780 |
| *Staal [11]* | 0.9440 | 0.7190 | 0.9770 | 0.9520 |
| *Mendonca[22]* | 0.9450 | 0.734 | 0.976 | - |
| *Zhang [23]* | 0.9476 | 0.7743 | 0.9725 | 0.9636 |
| *Li [24]* | 0.9527 | 0.7569 | 0.9816 | 0.9738 |
| *Dasgupta [25]* | 0.9533 | 0.7691 | 0.9801 | 0.9744 |
| *Yan [26]* | 0.9542 | 0.7653 | 0.9818 | 0.9752 |
| *Xie [27]* | 0.9560 | 0.7811 | 0.9839 | 0.9792 |
| *Ricci [28]* | 0.9590 | 0.7750 | 0.9630 | - |
| *Proposed* | **0.9700** | **0.7864** | **0.9876** | **0.9862** |

**Table 3** Comparison results on the STARE datasets

| Segment method | Acc | Sen | Spe | AUC |
|---|---|---|---|---|
| *Human observer* | 0.9339 | 0.8953 | 0.9374 | 0.9170 |
| *Staal [11]* | 0.9520 | 0.6970 | 0.9810 | 0.9610 |
| *Roychowdhury [20]* | 0.9560 | 0.7320 | 0.9840 | 0.9670 |
| *Zhao [29]* | 0.9560 | 0.7800 | 0.9780 | 0.8740 |
| *Li [24]* | 0.9628 | **0.7726** | 0.9844 | **0.9879** |
| *Ricci [28]* | 0.9584 | - | - | 0.9602 |
| *Soares [30]* | 0.9480 | 0.7200 | 0.9750 | 0.9670 |
| *Marin [31]* | 0.9526 | 0.6944 | 0.9819 | 0.9769 |
| *Fraz [32]* | 0.9534 | 0.7548 | 0.9763 | 0.9768 |
| *Lam [33]* | 0.9470 | - | - | 0.9740 |
| *Proposed* | **0.9680** | 0.6858 | **0.9893** | 0.9787 |

**Table 4** Comparison results on the HRF datasets

| Segment method | Acc | Sen | Spe | AUC |
|---|---|---|---|---|
| *Fraz [34]* | 0.9430 | 0.7152 | 0.9759 | - |
| *Odstrcilik [35]* | 0.9494 | 0.7741 | 0.9669 | - |
| *Yu [36]* | 0.9515 | **0.7811** | 0.9685 | - |
| *Annunziata [37]* | 0.9581 | 0.7128 | 0.9836 | - |
| *Proposed* | **0.9677** | 0.7249 | **0.9873** | **0.9827** |

## 4. CONCLUSION

In this paper, we propose a task-specific transfer method based on the information bottleneck theory to solve the data scarcity in medical image segmentation. Our method outperformed state-of-the-art methods with an accuracy of 97%, 96.8% and 96.77% on DRIVE, STARE and HRF respectively, outperforming both current competitor methods and independent human observers (DRIVE (96.37%) and STARE (93.39%)). Most importantly, we proved that with appropriate transfer methods, the selected cross-class databases may improve segmentation better than seemingly similar homogeneous databases.